\documentclass[twocolumn]{aastex631}
\usepackage{amsmath}
\usepackage{array}
\newcolumntype{?}{!{\vrule width 1.5pt}}
\newcommand{\minus}{\scalebox{0.75}[1.0]{$-$}}

\expandafter\ifx\csname natexlab\endcsname\relax\fi
\providecommand{\url}[1]{\href{#1}{#1}}
\providecommand{\dodoi}[1]{doi:~\href{http://doi.org/#1}{\nolinkurl{#1}}}
\providecommand{\doeprint}[1]{\href{http://ascl.net/#1}{\nolinkurl{http://ascl.net/#1}}}
\providecommand{\doarXiv}[1]{\href{https://arxiv.org/abs/#1}{\nolinkurl{https://arxiv.org/abs/#1}}}

\shorttitle{AT2021fxu}
\shortauthors{Ajay et al.}

\graphicspath{{./}{figures/}}

\begin{document}

\title{A Multiwavelength study of the ongoing Changing-look AGN AT2021fxu}

	\author{Yukta Ajay}
	\affiliation{Indian Institute of Science Education and Research (IISER), Tirupati\\
	Tirupati 517 507, Andhra Pradesh, India \\}
	
	    \author{Muryel Guolo}
    \affiliation{Department of Physics and Astronomy, Johns Hopkins University\\ 3400 N. Charles St., Baltimore MD 21218, USA}
    
	\author{Dheeraj Pasham}
	\affiliation{MIT Kavli Institute for Astrophysics and Space Research \\
		Cambridge, MA 02139, USA}

\begin{abstract}
 \noindent We present multiwavelength analyses of an active optical transient AT2021fxu which shows the appearance of previously absent broad emission lines in a recent optical spectrum, suggesting a Changing-look (CL) Active Galactic Nucleus (AGN). During the spectral transition, the brightness in the individual photometric bands increased up to $\approx$0.6 in the optical bands and up to $\approx$1.1 magnitudes in the UV bands. The brightening was accompanied by a blueward shift of the optical spectrum. AT2021fxu shows high X-ray (0.3-10 keV) flux variability before and after the outburst, with the average X-ray flux increasing by a factor of $\approx$2 post-outburst. However, the X-ray spectral shape remains roughly the same, with no significant change in the line-of-sight column density. AT2021fxu's overall properties are consistent with an accretion-rate-driven transition from a Type-II to a Type-1 AGN.
\end{abstract}

\section{Introduction} 
Accreting active galactic nuclei (AGN) can be identified by their optical emission line features such as broad and/or narrow high ionization lines. 
Doppler broadened Broad Emission Lines (BELs) originate from the gravitationally-bound broad-line region (BLR). In contrast, the narrow-line regions (NLRs) can extend beyond a few hundred parsecs, although the ionizing AGN continuum considerably influences the NLR. AGN that show both the broad- and the narrow-lines are classified as Type-I AGN, while those that do not show BELs are called as Type-II AGN. The AGN unification model \citep{1993ARA&A..31..473A, 1995PASP..107..803U} attributes differences in type to the viewing angle towards an axisymmetric, parsec scale toroidal structure (the so-called ``torus''). This structure can obstruct our direct view of the AGN central engine along our line-of sight, and block our view of the BLR in Type-II AGN. 

\par A small subset of AGN, referred to as changing-look (CL) AGN have been reported in recent years \citep[e.g.,][]{2015ApJ...800..144L, 2021MNRAS.508..144G, 2022ApJ...939L..16Z}. These  show strong spectral variability in their BELs and continuum, and can therefore appear to transition between the AGN classes over months to several years timescale. CL-AGN transitions have been explained mainly by variable obscuration or a strong variation in the accretion rate \citep[see][for a recent review]{2022arXiv221105132R}.

\par AT2021fxu/ZTF21aalxxzn/Gaia22dgm is an optical transient at a redshift of 0.1097 identified by the Zwicky Transient Facility (ZTF) on 13 February 2021 \citep{2021TNSTR.792....1D}. Prior to the optical outburst, it was cataloged as an AGN (2MASX J13471874+0210579/SWIFT J1347.7+0212) by the BAT AGN Spectroscopic Survey \cite[BASS,][]{2017ApJ...850...74K}. Here we present analysis of AT2021fxu's multiwavelength data taken over a temporal baseline of over two decades.

\section{Data and Analysis}
\subsection{Properties of optical spectra}
In Fig.\ref{fig:panel}a, we show AT2021fxu's optical spectra before and after the optical transition. The archival optical spectrum from Sloan Digital Sky Survey (SDSS) in 2001 shows a continuum dominated by stellar population with no clear BELs. Although a specific AGN class is hard to pin down due to contamination of sky lines in the H$\alpha$ region, the general features are consistent with either a low-ionization nuclear emission-line region \citep[LINER,][]{1980A&A....87..152H} or a Type-II AGN. In the optical spectrum captured by the Faulkes Telescope North in 2021 (Fig.\ref{fig:panel}c), a bright blue continuum and BELs typical for a Type-I AGN are evident. This indicates a change in the spectral class of the AGN from Type-II to Type-I.
\pagebreak
\subsection{Optical and UV photometry}
{\it Swift}'s Ultra-Violet/Optical
Telescope (UVOT) performed observations of the source before and after the optical outburst. The data were processed with the {\tt uvotsource} command
within the HEASoft v6.29c package. We also performed point spread function (PSF) photometry on all publicly available ZTF data at the location
of AT2021fxu using the ZTF forced-photometry service
\citep{2019PASP..131a8003M} in the g- and r-bands.
The measured magnitudes have been corrected for Galactic
extinction using E(B\minus V) = 0.021 from \citet{2011ApJ...737..103S}.
\par Fig.\ref{fig:panel}c shows the optical lightcurves since Decmeber 2020 ($\sim$MJD 59200), which is nearly flat until July 2021 ($\sim$MJD 59420). Additionally, the optical magnitudes recorded by SDSS in May 2000 (18.24$\pm$0.01 in the optical-g and 17.33$\pm$0.01 in the optical-r, not shown here) are consistent with the average optical magnitudes between March 2018 to July 2021 (18.22$\pm$0.08 (g-band) and 17.41$\pm$0.05 (r-band)). This suggests that AT2021fxu's optical spectrum did not change between 2000 and 2021. Further, the optical brightness of the source increased by $\approx$0.6 and $\approx$0.34 magnitudes in the g- and r-bands, respectively, during the optical outburst between 1 April 2022 and 27 May 2022. This was followed by an optical decline of $\approx$0.34 magnitudes in the g-band  and $\approx$0.2 magnitudes in the r-band over the next two months, until 19 July 2022. 
\par Comparing the average UV magnitudes from about a decade prior to the outburst with the values recorded during the optical decline, the magnitudes have changed from $\approx$18.3 to $\approx$17.7 in the U-band, from $\approx$19.3 to $\approx$18.2 in the UVW1 band, and from $\approx$19.6 to $\approx$18.4 in the UVW2 band, indicating that the brightness in all the UV bands increased post-outburst.  
\par There is a shift to a bluer spectral energy distribution (SED), with the optical (g\minus r) color decreasing from 0.80$\pm$0.06 to 0.46$\pm$0.04, before and after the optical outburst, respectively. At the optical peak, the (g-r) difference is minimum, and is $\approx$0.38. Similarly, the average UV (W1\minus U) magnitude difference drops from 1.02$\pm$0.07 to 0.5$\pm$0.01, and the average UV (W2\minus U) magnitude decreases from 1.28$\pm$0.05 to 0.74$\pm$0.02, again indicative of a bluer SED.

\subsection{X-ray spectral analysis}
The archival {\it Swift} X-Ray Telescope (XRT) observations between 4 December 2011 to 13 August 2012 suggest that the source was highly variable in X-rays, with the 0.3-10 keV count rate increasing by a factor of 10 in less than 2 months.
We fit the combined 0.3-10 keV XRT spectrum using the \texttt{TBabs*zTBabs*zashift(powerlaw)} model in XSPEC \citep{1996ASPC..101...17A} with the galactic column density ($N_H$) fixed at 1.5$\times$10$^{20}$  cm$^{-2}$ \citep{2016A&A...594A.116H}, and the host galaxy $N_H$ left as a free parameter. The best-fit powerlaw index and the neutral column density were $1.71_{-0.19}^{+0.42}$ and $<$$8\times10^{20}$ cm$^{-2}$, respectively. The observed luminosity and flux at these times were $2.72_{-0.64}^{+0.48}\times10^{43}$ ergs s\textsuperscript{-1} and $0.91\pm 0.20\times10^{-12}$ ergs cm\textsuperscript{-2} s\textsuperscript{-1}, respectively. All the errorbars reported in this work represent 90\% confidence values. 

\par {\it Swift}/XRT observations between August 2021 and August 2022 coincided with the recent optical outburst. We fit the combined XRT spectrum during the optical decline (mean countrate of $0.038\pm0.002$ cps) with the same powerlaw model as before and obtained a best-fit photon index and column denstiy values of $1.83_{-0.16}^{+0.32}$, and $<7.5\times10^{20}$ cm$^{-2}$, respectively. The observed 0.3-10 keV luminosity and flux of the post-outburst XRT spectrum were found to be $6.54_{-1.37}^{+0.76}\times10^{43}$ ergs s\textsuperscript{-1}  and $2.17_{-0.35}^{+0.31}\times10^{-12}$ ergs cm\textsuperscript{-2} s\textsuperscript{-1}, respectively.

\par
\begin{figure*}[ht!]
    \centering
    \includegraphics[scale=0.6]{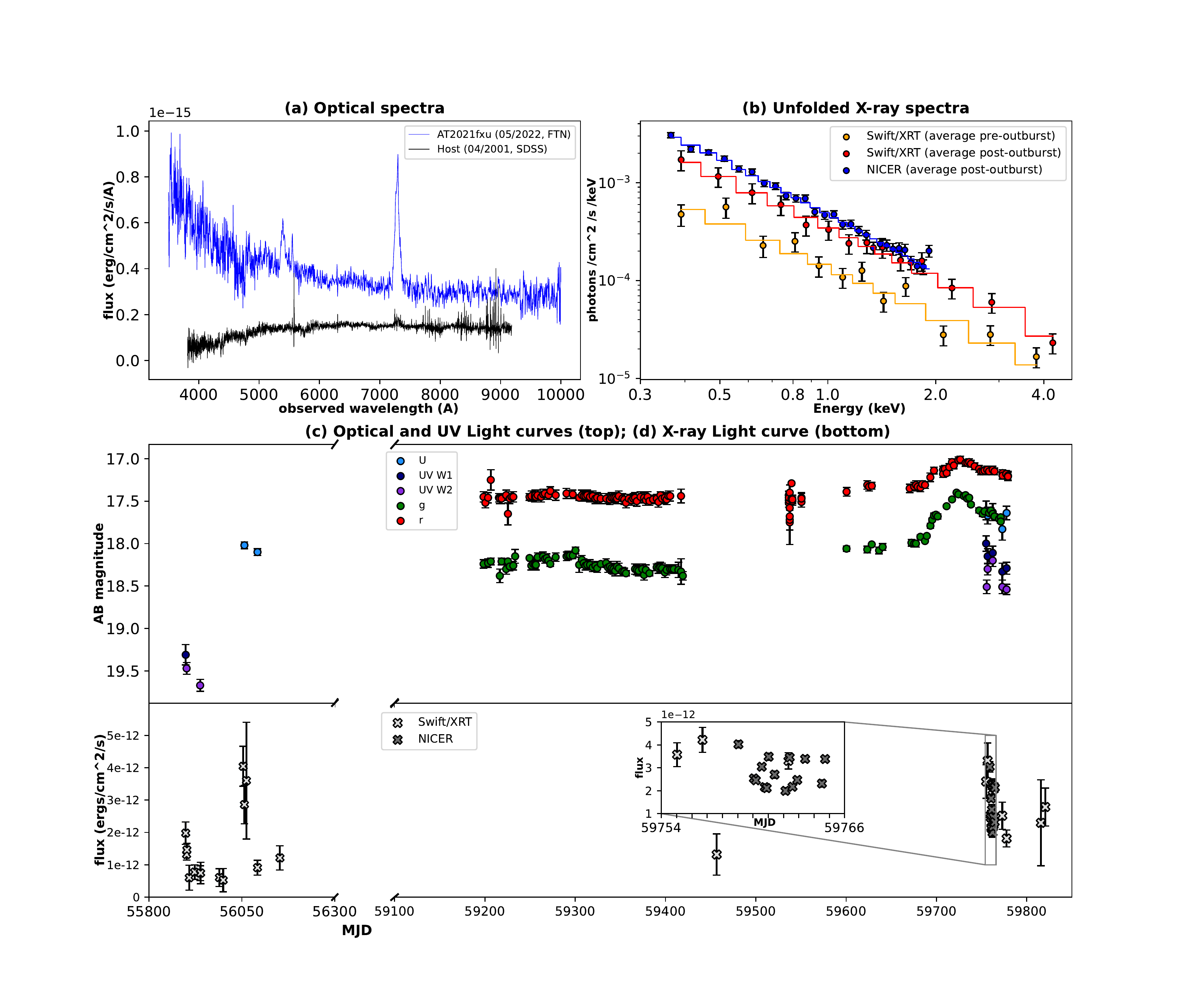}
    \caption{(a) Optical spectrum of the host galaxy in 2001 (black), and the optical spectrum post-outburst in May 2022 (blue) showing the appearance of BELs and a strong blue continuum; (b) Unfolded spectra and  models for {\it Swift}/XRT data between 2011-2012 (before the optical outburst), and {\it Swift}/XRT and {\it NICER} between June-August 2022 (post-outburst); (c) Optical and UV light curves from {\it Swift}/UVOT; (d) {\it Swift}/XRT and {\it NICER} X-ray light curves, zoomed in plot shows short-term X-ray variability.}
    \label{fig:panel}
\end{figure*}

 Following XRT’s detection, the Neutron Star Interior Composition Explorer{\it (NICER)} performed multiple exposures between 28 June 2022 and 4 July 2022 during the optical decline. The source was variable during this time (Fig.\ref{fig:panel}d) with a mean background-subtracted count rate of $1.33\pm0.02$ cps in the 0.3-2 keV band. We fit a combined {\it NICER} 0.3-2.0 keV spectrum using the powerlaw model described above. The best-fit power law index and the host $N_H$ column upper limits are $2.01^{+0.13}_{-0.06}$ and $1.5\times10^{20}$ cm$^{-2}$, respectively, consistent with the post-outburst values from {\it Swift}/XRT. The observed average luminosity and flux in the 0.3-10 keV energies from {\it NICER} are $7.98_{-0.82}^{+0.73}\times10^{43}$ ergs s\textsuperscript{-1} and $2.60_{-0.30}^{+0.25}\times10^{-12}$ ergs cm\textsuperscript{-2} s\textsuperscript{-1}, respectively. 

\par The X-ray lightcurves from {\it Swift}/XRT and {\it NICER} (Fig.\ref{fig:panel}d) show a doubling in the X-ray flux post-outburst and high flux variability with almost no change in the X-ray spectral shape over time. Furthermore, the powerlaw index and the host column density values do not show considerable variation between 2011-12 and 2021-22. 

\section{Discussion and Conclusion}
We have analyzed AT2021fxu's optical, UV and X-ray data over approximately two decades. The recent optical spectrum clearly indicates the appearance of BELs, suggesting that AT2021fxu underwent a spectral transition from a Type-II to a Type-I AGN. Since the optical g- and r- magnitudes have remained constant between May 2000 to July 2021, it is clear that optical spectrum remained the same during these years, and the recent optical outburst in April-May 2022 marks the onset of the CL transition in this AGN.
\par X-ray spectral analysis shows that there is minimal variation in the column density over the observed time. This eliminates the possibility of variable obscuration driving the spectral transition, implying instead that the CL transition is due to a variable accretion rate. This mechanism of transition due to increased accretion is further confirmed by an increase in the flux in all energy bands and a shift of the SED toward bluer wavelengths.
\par In summary, we show that AT2021fxu is an accretion-drive CL transition in the AGN 2MASX J13471874+021057/SWIFT J1347.7+0212, where the increase in the ionization flux is able to produce strong broad Balmer lines that drives the change in spectral classification.
\par Interestingly, during the writing of this note, the source showed further optical brightening by $\approx$0.12 magnitudes. We have initiated follow-up observations on {\it Swift} and {\it NICER} to further analyze this new outburst, and we encourage observations in other wavelengths to study the ongoing CL transition.

\newpage


\end{document}